\documentclass[aps,prl,twocolumn,showpacs,superscriptaddress,groupedaddress]{revtex4}  
\usepackage{graphicx}  
\usepackage{dcolumn}   
\usepackage{bm}        
\usepackage{amssymb}   
\usepackage{color}

\begin{document}


\title{Localization of charge carriers in monolayer graphene
gradually disordered by ion irradiation}

\author{E. Zion}
\affiliation{Department of Physics and Institute of Nanotechnology and Advanced Materials, Bar-Ilan University, Ramat-Gan 52900, Israel}

\author{A.Haran}
\altaffiliation[Permanent address: ]
{Soreq NRC, Yavne 8180000, Israel}
\affiliation{Faculty of Engineering, Bar-Ilan University, Ramat-Gan 52900, Israel}

\author{A. V. Butenko}
\affiliation{Department of Physics and Institute of Nanotechnology and Advanced Materials, Bar-Ilan University, Ramat-Gan 52900, Israel}

\author{L. Wolfson}
\affiliation{Jack and Pearl Resnick Institute, Department of Physics, Bar-Ilan University, Ramat-Gan 52900, Israel}

\author{Yu. Kaganovskii}
\affiliation{Jack and Pearl Resnick Institute, Department of Physics, Bar-Ilan University, Ramat-Gan 52900, Israel}

\author{T. Havdala}
\affiliation{Department of Physics and Institute of Nanotechnology and Advanced Materials, Bar-Ilan University, Ramat-Gan 52900, Israel}

\author{A. Sharoni}
\affiliation{Department of Physics and Institute of Nanotechnology and Advanced Materials, Bar-Ilan University, Ramat-Gan 52900, Israel}

\author{D. Naveh}
\affiliation{Faculty of Engineering, Bar-Ilan University, Ramat-Gan 52900, Israel}

\author{V. Richter}
\affiliation{Solid State Institute and Physics Department, Technion-Israel Institute of Technology, Haifa 32000, Israel}

\author{M. Kaveh}
\affiliation{Jack and Pearl Resnick Institute, Department of Physics, Bar-Ilan University, Ramat-Gan 52900,
Israel}

\author{E. Kogan}
\affiliation{Jack and Pearl Resnick Institute, Department of Physics, Bar-Ilan University, Ramat-Gan 52900, Israel}

\author{I. Shlimak}
\email{Issai.Shlimak@biu.ac.il}
\affiliation{Jack and Pearl Resnick Institute, Department of Physics, Bar-Ilan University, Ramat-Gan 52900, Israel}

\date{\today}

\begin{abstract}
Gradual localization of charge carriers was studied in a series of micro-size samples of monolayer graphene fabricated on the common large scale film and irradiated by different doses of C$^+$ ions with energy 35 keV.
Measurements of the temperature dependence of conductivity and magnetoresistance in fields up to 4 T showed that at low  disorder, the samples are in the regime of weak localization and antilocalization. Further increase of disorder leads to strong localization regime, when conductivity is described by the variable-range-hopping (VRH) mechanism. A crossover from the Mott regime to the Efros-Shklovskii regime of VRH is observed with decreasing temperature. Theoretical analysis of conductivity in both regimes  showed a remarkably good agreement with experimental data.
\end{abstract}

\pacs{73.22.Pr}

\maketitle

\section{Introduction}

Graphene, a sheet of sp$^2$
-bonded carbon atoms, for several years already continues to be in the focus of attention of physics community, due to its potentially transformative
impact across a wide range of applications including
advanced electronics and sensing. Carrier scattering in graphene can be due to all kinds of disorder,
including ripples in the graphene layer,
point defects and their associated short-range potentials,
charged impurities
residing in the supporting substrate, and adsorbed atoms on
the surface.
Investigation of the influence of disorder on the properties of graphene is attracting a tremendous interest due to possibility to modify this novel and promising material using weak or strong localization of charge carriers. By controllably introducing
defects into graphene, one may be able to understand
how these mechanisms limit transport.

Previosly, there were  observed separately either weak localization (WL)  or different kinds of Variable-Range-Hopping (VRH) conductivity of strongly localized carriers in graphene samples disordered by different methods like doping, oxidation, ion irradiation (see, for example, \cite{morosov,wu,tikhon,chen,lundeberger,buchowicz,jobst,moktadir}). However, we are not aware of observations of all regimes of localization with gradual increase of disordering in graphene. In this paper we report the results of  study of the localization process in monolayer graphene (MG) samples subjected by different doses of ion irradiation.

The initial large size ($5\times 5$ mm) specimens were supplied by "Graphenea" company. Monolayer graphene was produced by CVD on copper catalyst and transferred to a 300 nm SiO$_2$/Si substrate using wet transfer process. It is specified in the certificate, that the sheet resistivity of the specimen is 350 Ohm/sq. On one specimen, gold electrical contacts were deposited directly on the graphene surface. This sample is marked as 0. The resistivity of the sample 0 was 380 Ohm/sq which is close to the data in the certificate. On the surface of the other specimen, six groups of mini-samples ($200\times 200$ $\mu$m) were fabricated by means of electron-beam lithography (EBL) as well as electrical contacts (5 nm Ti and 45 nm Pd) for 2-probe measurements. The samples from the first group, marked as sample 1 were not irradiated, while 5 other groups were subjected to different doses (from $5\times 10{^13}$ up to $1\times 10^{15}$ cm$^{-2}$) of irradiation by C$^+$ ions with energy 35 keV. Ion irradiation was performed on HVEE-350 Implanter, C$^+$ ions were obtained from the hollow cathode ion source by CO$_2$ decomposition.

The choice of the irradiation conditions was not occasional. They are the same as in the Ref. \cite{buchowicz} where the Hall effect was measured in monolayer graphene subjected to ion irradiation. Resistances of the initial sample in Ref. \cite{buchowicz} and our initial sample 0 are the same. It was shown in \cite{buchowicz} that both Hall coefficient and mobility are temperature independent up to high dose of irradiation. Moreover, it was shown that the sheet Hall concentration of charge carriers (about $10^{13}$ cm$^{-2}$) does not depend at all on the dose of irradiation. We were based on these results in our decision to prefer the 2-probe geometry instead of the Hall bar geometry, because this allows us to measure more samples on the same sample holder. As a result, we were able to enhance the reproducibility of the obtained results. Comparison of the graphene resistance measured by 2-probe method and by 4-probe method presented in \cite{buchowicz} showed that in this system, contact resistance is insignificant.

\begin{figure}[h]
\includegraphics[width= .9\columnwidth]{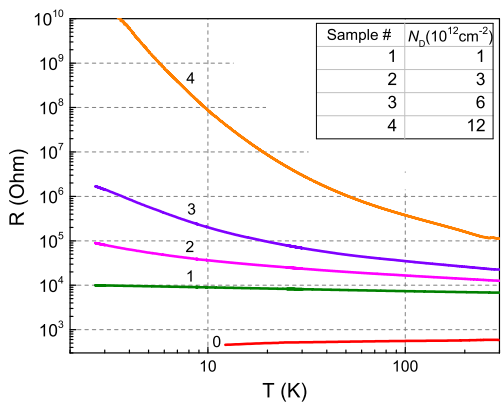}
\caption{\label{fig:log}  Resistivity of disordered monolayer graphene samples as a function of temperature. Inset shows  the density of structural defects in samples \cite{shlimak}. }
\end{figure}

In our previous work \cite{shlimak}, a concentration of structural defects $N_D$ was determined for each group of samples using measurements of the Raman scattering. These values are shown in the inset in Fig. \ref{fig:log}. (It turns out that sample 1 is also slightly disordered due to EBL process). Measurements of the current-voltage characteristics ($I-V$ ) for all samples, performed at room temperature, showed that for samples 5 and 6, $I-V$ is strongly non-linear even at very small current. That is why in this paper, the temperature dependences of resistance R(T) are shown only for samples 0 - 4.

The resistance was measured by two-probe method in  helium cryostat down to 1.8 K in zero magnetic field and in magnetic fields up to 4 Tesla. Fig. \ref{fig:log} shows the general picture $R(T)$ for all samples.

The sample 0 shows typical metallic behavior, when $R$ slightly decreases with decrease of $T$. For sample 1 $R$ slightly increases with decreasing
$T$, which is characteristic for "dirty" metals.  For other samples $R$ changes with $T$  exponentially, which is characteristic for strongly localized carriers.

\section{Weak localization}

\begin{figure}[h]
\includegraphics[width= .8\columnwidth]{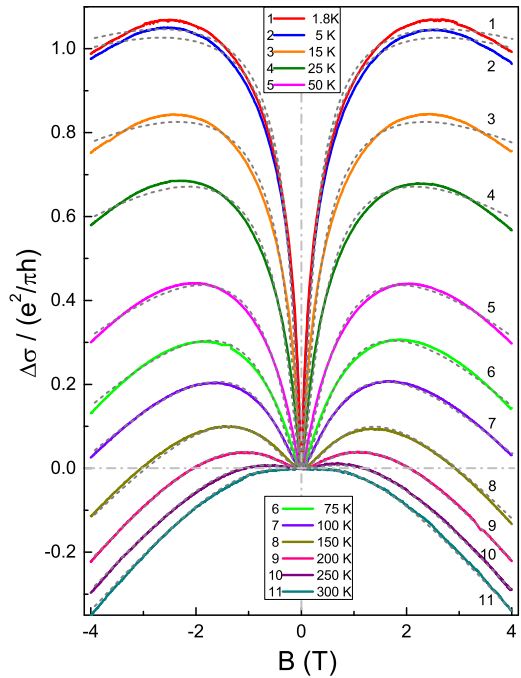}
\caption{\label{fig:fitting}  Magnetoconductance of sample 1  as function of magnetic field;  solid lines -- experiment, dashed lines --  formula  (\ref{phenomenological}) with fitted parameters. }
\end{figure}

\begin{figure}[h]
\hspace*{-.75 cm}
\includegraphics[width= 1.2\columnwidth]{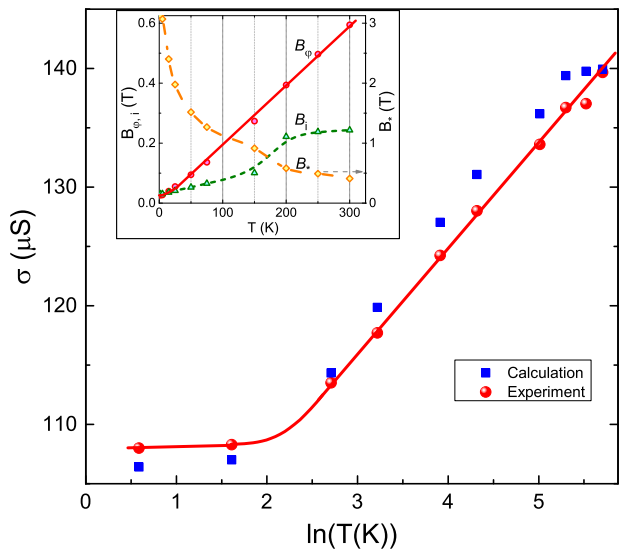}
\caption{\label{fig:sigma1}  Conductivity  of sample 1 as function of temperature. Circles present experimental data, squares present Eq. (\ref{alt2}) with parameters determined from fitting the magnetoconductance
($A$ was chosen to be $11.7\cdot e^2/\pi h$).
The inset: The values  of $B_{\varphi}$  (solid line) and $B_i$ (short-dashed line) and $B_*$ (long-dashed line, right axis).}
\end{figure}

Fig. \ref{fig:fitting} shows the experimental dependences of magnetoconductance (MC) of sample 1 in wide temperature interval, from 300 K down to 1.8 K. Plot of the temepature dependence of conductivity on the scale $\sigma$ vs. $\ln T$ (Fig. \ref{fig:sigma1})  shows  the logarithmic temperature behavior  of $\sigma$ at low $T$, characteristic for regime of WL \cite{khmel}, with tendency to saturation at very low temperatures.

WL regime of conductivity in monolayer graphene has  important features due to the facts that charge carriers are chiral Dirac fermions, which are reside in two inequivalent valleys at the $K$ and $K'$ points of the  Brillouin zone. Due to chirality, Dirac fermion  acquires a phase of $\pi$ upon intravalley
scattering,  which leads to destructive interference with its time-reversed counterpart and weak antilocalization (WAL). Intervalley scattering leads to restoration of WL because fermions in $K$ and $K'$ valleys have opposite chiralities.

Quantum corrections to the conductivity of graphene have been intensely studied theoretically
\cite{morpurgo,aleiner,altland,ostrovsky,cann,kechedzhi,tikhonenko,pal,baker}. It was predicted that at relatively high temperatures WAL corrections will dominate, while with decreasing $T$ the WL corrections will dominate.
There were  several experimental papers reporting logarithmic dependence of conductivity on temperature and magnetic field  at low temperatures \cite{morosov,wu,tikhon,chen,lundeberger,jobst}.
However, in our sample 1, the logarithmic dependence is observed in wide temperature interval, starting from 300 K, which gives an opportunity to check in a very detailed way the theoretical predictions.

 For the MC
the theory \cite{cann} predicts
\begin{eqnarray}
\label{phenomenological}
&&\Delta \sigma(B,T)=\frac{e^2}{\pi h}\left[F\left(\frac{B}{B_{\varphi}}\right)-F\left(\frac{B}{B_{\varphi}+2B_i}\right)\right.\nonumber\\
&&\left.-2F\left(\frac{B}{B_{\varphi}+B_*}\right)\right]\\
&&F(z)=\ln(z)+\psi\left(\frac{1}{2}+\frac{1}{z}\right),\;\;\;B_{\varphi,i,*}=\frac{\hbar c}{4De} \tau_{\varphi,i,*}^{-1}
\nonumber,
\end{eqnarray}
where $\psi$ is the digamma function,  $\tau_{\varphi}$ is the coherence time, $\tau_i^{-1}$ is the intervalley scattering rate, $\tau_*^{-1}$ is
the combined scattering rate of intravalley and intervalley scattering and of trigonal warping.

Fitting  Eq. (\ref{phenomenological}) to experimental data for magnetoconductance of sample 1 at different temperatures   is illustrated on Fig. \ref{fig:fitting}.
In the process of fitting   we were able to extract all three parameters  $B_{\varphi,i,*}$ entering the equation.
These parameters shown in inset in Fig. \ref{fig:sigma1}. It turns out, that they are temperature--dependent which was not predicted by theory.
Using these parameters we can calculate conductance at zero magnetic field according to Eq. (10) of Ref. \cite{cann}, which can be rewritten in the form
\begin{eqnarray}
\label{alt2}
\sigma(B=0,T)&=&-\frac{e^2}{\pi h}\left[\ln\left(1+2\frac{B_i}{B_{\varphi}}\right)+2\ln\left(1+\frac{B_*}{B_{\varphi}}\right)\right.\nonumber\\
&&\left.+2\ln \left(\frac{B_{\varphi}}{1\;{\rm T}}\right)\right]+A,
\end{eqnarray}
where   $A$ is a constant, dependent upon the unit of magnetic field (chosen as  1 T).
We compare Eq. (\ref{alt2})  with the experimentally measured
conductivity  $\sigma(B=0,T)$.  The comparison is presented on Fig. \ref{fig:sigma1}. The good agreement  proves the correctness of the chosen parameters.

The inset to Fig. \ref{fig:sigma1} shows that, apart from the lowest temperatures,
$B_{\varphi}\sim 1/\tau_{\varphi}\sim T$.
Mechanisms that can give  the dependence $\tau_{\varphi}\sim 1/T$ are: electron--electron scattering in dirty limit \cite{khmel}, electron--phonon scattering,
\cite{stauber,hwang},
 and electron-flexural phonon  interaction
\cite{fink}. The saturation of $\tau_{\varphi}$ at low temperatures is well known in classical 2d systems and may be connected with existence of dephasing centers (for example, magnetic impurities) \cite{bishop,gantmakher}.

Obtained value of $B_{\varphi}$ allow us to determine the values of dephasing length
$L_{\varphi}=\sqrt{D\tau_{\varphi}}=\sqrt{\hbar /4B_{\varphi}e}$.
When the temperature  decreases from 300 K to 3 K, the dephasing length $L_{\varphi}$  increases from 7 nm to 70 nm and then saturates.
The maximal value of $L_{\varphi}$ allows us to estimate the density of dephasing centers as $2\times 10^{10}$ cm$^{-2}$.


\section{Strong localization}

Let's discuss now samples 2-4 with pronounced insulating behaviour. Plotting the data on the Arrhenius scale  $\ln R$ vs. $1/T$ showed that energy of activation continuously decreases with decreasing $T$ which is characteristic for the variable-range-hopping (VRH) conductivity \cite{shklovskii}. There are two kinds of VRH depending on the structure of the density-of-states (DOS)  $g(\epsilon)$ in the vicinity of the Fermi level (FL) $\mu$: when $g(\epsilon) = g(\mu)=$const, $R(T)$ is described by the Mott ($T^{-1/3}$) law in the case of two-dimensional (2d) conductivity:
\begin{eqnarray}
\label{5}
R(T) = R_0 \exp(T_M/T)^{1/3},\;\;\;  	T_M = C_M[g(\mu)a^2]^{-1}.   		
\end{eqnarray}
Here $C_M = 13.8$ is the numerical coefficient \cite{shklovskii}, $a$ is the radius of localization.

The Coulomb interaction between localized carriers leads to appearance of the soft Coulomb gap in the vicinity of FL which in the case of 2d has a linear form
\begin{eqnarray}
\label{6}
g(\epsilon)\sim |\epsilon-\mu|(e^2/\kappa)^{-2},		
\end{eqnarray}
where $\kappa$ is the  dielectric constant of the material. This leads to the  Efros-Shklovskii (ES) VRH or $T^{-1/2}$ law:
\begin{eqnarray}
\label{7}
R(T) = R_0 \exp(T_{ES}/T)^{1/2}, \;\;\; 	T_{ES} = C_{ES} (e^2/\kappa a),	
\end{eqnarray}
where the numerical coefficient $C_{ES}= 2.8$ \cite{shklovskii}.

\begin{figure}[h!]
\vskip -2cm
\includegraphics[width=.83\columnwidth]{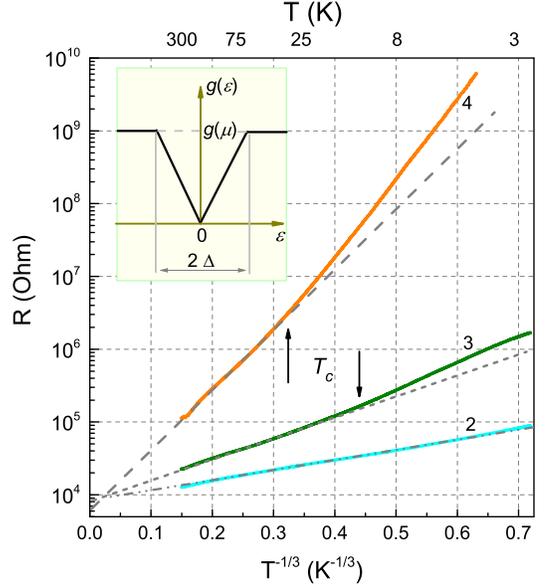}
\vskip -1cm
\caption{\label{Fig7}$\log R$ for samples 2-4  plotted versus $T^{-1/3}$.}
\end{figure}

Coulomb interaction can alter the DOS only near the FL. Far from FL, the DOS is restored to its initial value, which is approximately equal to
$g(\mu)$, see inset in Fig. \ref{Fig7}. Denoting the half-width of the Coulomb gap as $\Delta$ one can conclude, therefore, that $T \ll \Delta$,  ES law has to be observed, while in the opposite case ($T \gg \Delta$), the Mott law should dominate.

\begin{figure}[h!]
\vskip -1.7cm
\includegraphics[width=.8\columnwidth]{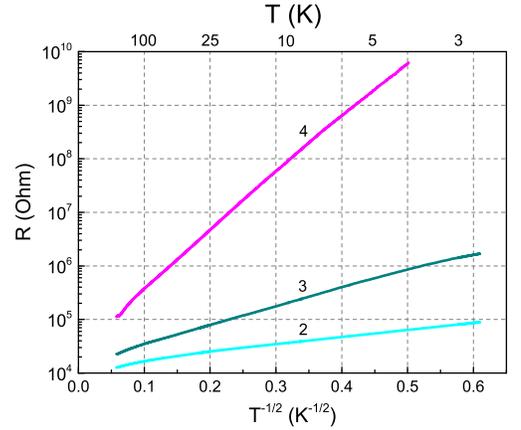}
\vskip -2.7cm
\caption{\label{Fig8}$\log R$ for samples 2-4  plotted versus  $T^{-1/2}$.}
\end{figure}

There is a number of reports about observation of either Mott or ES  laws in different disordered graphene-based materials \cite{moser,joung,hong,zhang}. We show that in samples 3 and 4, both VRH laws are observable at different temperatures. (For sample 2, the VRH regime will be observed at lower temperatures). In Figs. \ref{Fig7} and \ref{Fig8}, $\log R$ is plotted versus $T^{-1/3}$ and $T^{-1/2}$. At high temperatures, dependences $R(T)$ are straightened on the scale $T^{-1/3}$, while at low temperatures they are straightened on the scale $T^{-1/2}$. The latter shows the approach to the ES law which should be observed at the lowest temperatures. These plots allow us to determine both parameters $T_M$ and $T_{ES}$ (Table 1) and calculate the temperature $T_c$ of deviation from $T^{1/3}$ law to $T^{1/2}$ law in the case of 2d conductivity similarly to the calculation of $T_c$ for 3d conductivity \cite{shlimak2}.
\begin{table}
\begin{tabular}{|c|c|c|c|c|}
\hline
S $\#$  & $T_M$(K) & $T_{ES}(K)$ &  $T_c$(K) & $\Delta$(K)  \\
\hline
3 & $308\pm 54$ & $50\pm 7$ &  $11.4\pm 0.8$ & $12\pm 1.2$  \\
\hline
4 & $5962\pm 305$ & $490\pm 12$ &  $29\pm 5$ & $60\pm 6$  \\
\hline
\end{tabular}
\caption{Hopping conductivity parameters for samples 3,4.}
\end{table}
In VRH, only localized states  in an optimal band of width $\epsilon(T)$  near the Fermi level  are involved in the hopping process. The band becomes continuously narrower with decreasing $T$. Hopping resistance is determined by the critical parameter $\xi_c$:
\begin{eqnarray}
\label{8}
R = R_0 \exp \xi_c,\;\;\; 		\xi_c = \left(2r/a \right) + \left(\epsilon/T \right).	
\end{eqnarray}
Here energy and temperature are measured in the same units, $r$ is the mean distance of hopping. In the Mott regime, $g(\epsilon) = g(\mu) =$ const and, therefore, the total number of
states in the optimal band is $N(T) = g(\mu)\epsilon$ and the mean distance between states in two dimensions is
$r \approx [g(\mu)\epsilon]^{-1/2}$. Substituting  in (\ref{8}), one can find $\epsilon(T)$  from the minimal value of $\xi_c$ determined from
$d\xi_c/d\epsilon = 0$:
\begin{eqnarray}
\label{9}
\epsilon(T) = T^{2/3} [g(\mu)a^2]^{-1/3}.			
\end{eqnarray}
This gives the  relationship between $T$ and the width of the optimal band:
$T = [g(\mu)a^2]^{1/2}\epsilon^{3/2}$. At the crossover temperature $T_c$,  $\epsilon = \Delta$, so
\begin{eqnarray}
\label{10}
T_c = [g(\mu)a^2]^{1/2} \Delta^{3/2}.				
\end{eqnarray}
Being inside the Coulomb gap, the crossover temperature can be  determined from $g(\epsilon=\Delta) = g(\mu)$ which gives
$\Delta = g(\mu)(e^2/\kappa)^2$. Substituting  into Eq. (\ref{10}) and using expressions for $T_M$ and $T_{ES}$, Eqs. (\ref{5}) and (\ref{7}), we get
\begin{eqnarray}
\label{11}
Tc = (C_M^2/C_{ES}^3) (T_{ES}^3/T_M^2)\approx 8.6(T_{ES}^3/T_M^2)	
\end{eqnarray}
The functional proportionality of $T_c$ to the ratio ($T_{ES}^3/T_M^2$) has been obtained earlier in Ref. \cite{singh,lien}, but with significantly different numerical coefficient. The latter is, however, crucial for comparison with experiment. The values of  $T_c$ calculated from Eq. (\ref{11}) for samples 3 and 4 are given in Table 1 and shown as arrows in Fig. \ref{Fig7}. The good agreement shows the correctness of the obtained numerical coefficient.  We can also estimate the width of the Coulomb gap. In calculation, the value of $\kappa = 2.45$ for the monolayer graphene on the SiO$_2$ surface was used as $\kappa = (\kappa_1 + \kappa_2)/2$, where $\kappa_1 = 3.9$ (for SiO$_2$) and
 $\kappa_2 = 1$ (for air).
One can see that indeed, the ES law is observed when $T < \Delta$.

Comparison of samples 3 and 4 shows that increase of the density of defects $N_D$ leads to stronger localization which manifests in a significant decrease of $a$ and increase of the energy band needed for hopping. This looks like increase of the amplitude of the random potential relief in classical semiconductors induced by randomly distributed  positively charged donors and negatively charged acceptors in the case of compensation. Hence we assume that structural defects in graphene are of amphoteric impurity action \cite{dunlap}, i.e. they can be either acceptors or donors and compensate each other. Thus, increase of $N_D$  leads to the random potential relief  amplitude increase.
One can also assume that this phenomena may be connected with the fact that some point defects (say, vacancies) may produce complex associative centers with other defects. These associations can show an amphoteric impurity action, whereas the individual components are not amphoteric  \cite{fistul}.

In conclusion, a gradual transformation of conductivity measured in a wide interval of temperatures (300 - 1.8 K) and magnetic fields (up to 4 T) was observed in a series of monolayer graphene samples subjected by different dose of ion irradiation. Increasing the density of structural defects $N_D$ induced by irradiation has led to change the mechanism of electron transport from metallic conductivity to the regime of WL and finally to the VRH  of strongly localized carriers. It is shown that WL regime in slightly disordered sample starts from high (room) temperatures.

Comparison of experimental magnetoconductance curves with theory allowed us to find the parameters which determine the logarithmic temperature dependence of conductivity in WL regime. In VRH, a crossover from the Mott law to the Efros-Shklovskii law was observed in the same samples with decreasing temperature. The calculated crossover temperatures are in good agreement with experimental values. It is suggested that strengthening of localization with increase of $N_D$ can be explained by amphoteric impurity action of graphene structural defects induced by ion irradiation.


\begin{thebibliography}{99}

\bibitem{morosov} S. V. Morozov, K. S. Novoselov, M. I. Katsnelson, F. Schedin, L. A. Ponomarenko, D. Jiang, and A. K. Geim, Phys. Rev. Lett. {\bf 97},
016801 (2006).

\bibitem{wu} Xiaosong Wu, Xuebin Li, Zhimin Song, Claire Berger, and Walt A. de Heer, Phys. Rev. Lett. {\bf 98}, 136801 (2007).

\bibitem{tikhon} F. V. Tikhonenko, A. A. Kozikov, A. K. Savchenko, and R. V. Gorbachev, Phys. Rev. Lett. {\bf 103}, 226801 (2009).

\bibitem{chen} Yung-Fu Chen, Myung-Ho Bae, Cesar Chialvo, Travis Dirks, Alexey Bezryadin and Nadya Mason, J. Phys.: Condens. Matter {\bf 22}, 205301 (2010).

\bibitem{lundeberger} M. B. Lundeberg and J. A. Folk, Phys. Rev. Lett. {\bf 105},
146804 (2010).

\bibitem{buchowicz} G. Buchowicz, P.R. Stone, J.T. Robinson, C.D. Cress, J.W. Beeman, O.D. Dubon, Appl. Phys. Lett. {\bf 98}, 032102 (2011).

\bibitem{jobst} Johannes Jobst, Daniel Waldmann, Igor V. Gornyi, Alexander D. Mirlin, and Heiko B. Weber, Phys. Rev. Lett. {\bf 108}, 106601 (2012).

\bibitem{moktadir}Zakaria Moktadir, Shuojin Hang, Hiroshi Mizuta, arXiv:1410.4400.

\bibitem{shlimak}
I. Shlimak, A. Haran, E. Zion, T. Havdala, Yu. Kaganovskii, A. V. Butenko, L. Wolfson, V. Richter, D. Naveh, A. Sharoni, E. Kogan, and M. Kaveh, Phys. Rev. B {\bf 91}, 045414 (2015).

\bibitem{khmel} B. L. Altshuler A. G. Aronov, and D. E. and Khmelnitsky, J.
Phys. C: Solid State Phys. {\bf 15}, 7367 (1982).



\bibitem{morpurgo} A. F. Morpurgo and F. Guinea, Phys. Rev. Lett. {\bf 97}, 196804
(2006).

\bibitem{aleiner} I. L. Aleiner and K. B. Efetov, Phys. Rev. Lett. {\bf 97}, 236801 (2006).

\bibitem{altland} A. Altland, Phys. Rev. Lett. {\bf 97}, 236802 (2006).

\bibitem{ostrovsky} P. M. Ostrovsky, I. V. Gornyi, and A. D. Mirlin, Phys. Rev. B
{\bf 74}, 235443 (2006).

\bibitem{cann}E. McCann et al.,
Phys. Rev. Lett. {\bf 97} 146805 (2006).

\bibitem{kechedzhi} K. Kechedzhi, E. McCann, V. I. Fal'ko, H. Suzuura,
T. Ando, and B. L. Altshuler, The European Physics Journal
{\bf 148}, 39 (2007).

\bibitem{tikhonenko} F. V. Tikhonenko, A. A. Kozikov, A. K. Savchenko, and
R. V. Gorbachev, Phys. Rev. Lett. {\bf 103}, 226801 (2009).

\bibitem{pal} A. N. Pal, V. Kochat, and A. Ghosh, Phys. Rev. Lett.
{\bf 109}, 196601 (2012).

\bibitem{baker} A. M. R. Baker, J. A. Alexander-Webber, T. Altebaeumer,
T. J. B. M. Janssen, A. Tzalenchuk, S. Lara-Avila, S. Kubatkin,
R. Yakimova, C.-T. Lin, L.-J. Li, and R. J.
Nicholas, Phys. Rev. B {\bf 86}, 235441 (2012).

\bibitem{stauber} T. Stauber, N. M. R. Peres, and F. Guinea, Phys. Rev.
B {\bf 76}, 205423 (2007).

\bibitem{hwang} E. H. Hwang and S. Das Sarma, Phys. Rev. B {\bf 77}, 115449
(2008).

\bibitem{fink} K. S. Tikhonov, W. L. Z. Zhao, and A. M. Finkelstein, Phys. Rev. Lett. {\bf 113}, 076601 (2014).

\bibitem{bishop} D. J. Bishop, D. C. Tsui, R. C. Dines, Phys. Rev. Lett. {\bf 44}, 1153 (1980).

\bibitem{gantmakher} V.F. Gantmakher, {\it Electrons and Disorder in Solids}, (Oxford University Press, 2005).

\bibitem{shklovskii} B.I. Shklovskii and A.L. Efros, {\it Electronic Properties of Doped Semiconductors} (Springer-Verlag, Berlin, 1984).

\bibitem{moser}J. Moser, H. Tao, S. Roche, F. Alzina, C. M. Sotomayor Torres, and A. Bachtold,
Phys. Rev. B {\bf 81},  205445 (2010).

\bibitem{joung} D. Joung and S. Khondaker, Phys. Rev. B {\bf 86}, 235423 (2012).

\bibitem{hong} X. Hong, S.-H. Cheng, C. Herding, and J. Zhu, Phys. Rev. B {\bf 83}, 085410 (2011).

\bibitem{zhang} Haijing Zhang, Jianming Lu, Wu Shi, Zhe Wang, Ting Zhang, Mingyuan Sun, Yuan Zheng, Qihong Chen, Ning Wang, Juhn-Jong Lin, and Ping Sheng, Phys. Rev. Lett. {\bf 110}, 066805 (2013).

\bibitem{shlimak2} I. Shlimak, M. Kaveh, M. Yosefin, M. Lea and P. Fozooni, Phys. Rev. Lett. {\bf 68}, 3076 (1992)

\bibitem{singh} M. Singh, Y. Tarutani, U. Kabasava, and K. Takagi, Phys. Rev. B, {\bf 50}, 7007 (1994).

\bibitem{lien} Nguen V. Lien, Physics Letters A {\bf 207}, 379 (1995); Nguen V. Lien and R. Rosenbaum, Phys. Rev. B {\bf 56}, 14960 (1997).

\bibitem{dunlap}W. Crawford Dunlap, Phys. Rev. 100, 1629 (1955).

\bibitem{fistul} V. Fistul, {\it Impurities in Semiconductors: Solubility, Migration and Interactions}, (CRS Press, 2004).

\end{thebibliography}
\end{document}